\documentclass[aps, prb, twocolumn,groupedaddress]{revtex4} 
\usepackage{graphicx}
\usepackage{amsmath}
\usepackage{amssymb}
\usepackage{ifthen}
\usepackage{booktabs}

\newcommand{\bb}{\begin{equation}}
\newcommand{\ee}{\end{equation}}
\newcommand{\ba}{\begin{eqnarray*}}
\newcommand{\ea}{\end{eqnarray*}}
\newcommand{\rhor}{\rho({\bf r})}
\newcommand{\dd}{{\rm d}}
\newcommand{\rr}{{\mathbf r}}
\newcommand{\dr}{{\rm d}{\bf r}}

\usepackage{graphicx}
\usepackage{dcolumn}
\usepackage{bm}
\usepackage{longtable}

\begin{document}



\title{Condensation and evaporation transitions in deep capillary grooves}

\author{Alexandr \surname{Malijevsk\'y}}
\affiliation{ {Department of Physical Chemistry, ICT Prague, 166 28 Praha 6, Czech Republic} {ICPF, Academy of Sciences, 16502 Prague 6, Czech Republic}}
\author{Andrew O. \surname{Parry}}
\affiliation{Department of Mathematics, Imperial College London, London SW7 2B7, UK}

\begin{abstract}
We study the order of capillary condensation and evaporation transitions of a simple fluid adsorbed in a deep capillary groove using a fundamental measure
density functional theory (DFT). The walls of the capillary interact with the fluid particles via long-ranged, dispersion, forces while the fluid-fluid
interaction is modelled as a truncated Lennard-Jones-like potential. We find that below the wetting temperature $T_w$ condensation is first-order and evaporation
is continuous with the metastability of the condensation being well described by the complementary Kelvin equation. In contrast above $T_w$ both phase
transitions are continuous and their critical singularities are determined. In addition we show that for the evaporation transition above $T_w$ there is an
elegant mapping, or covariance, with the complete wetting transition occurring at a planar wall. Our numerical DFT studies are complemented by analytical slab
model calculations which explain how the asymmetry between condensation and evaporation arises out of the combination of long-ranged forces and substrate
geometry.
\end{abstract}


\maketitle

\section{Introduction}

\begin{figure*}
\includegraphics[width=5cm]{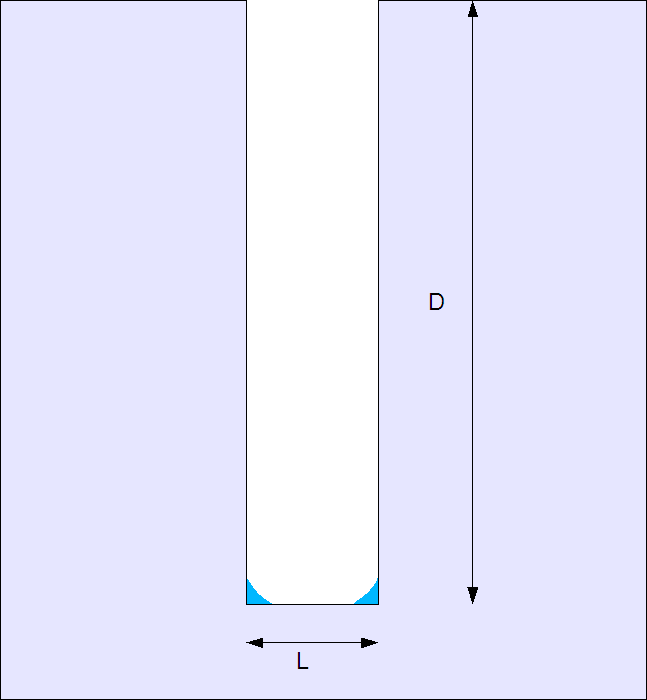}  \hspace*{0.5cm} \includegraphics[width=5cm]{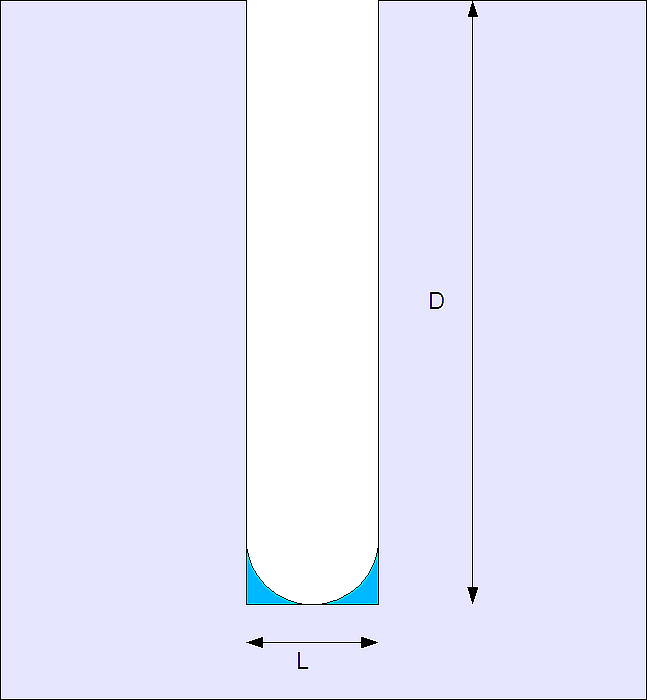}  \hspace*{0.5cm} \includegraphics[width=5cm]{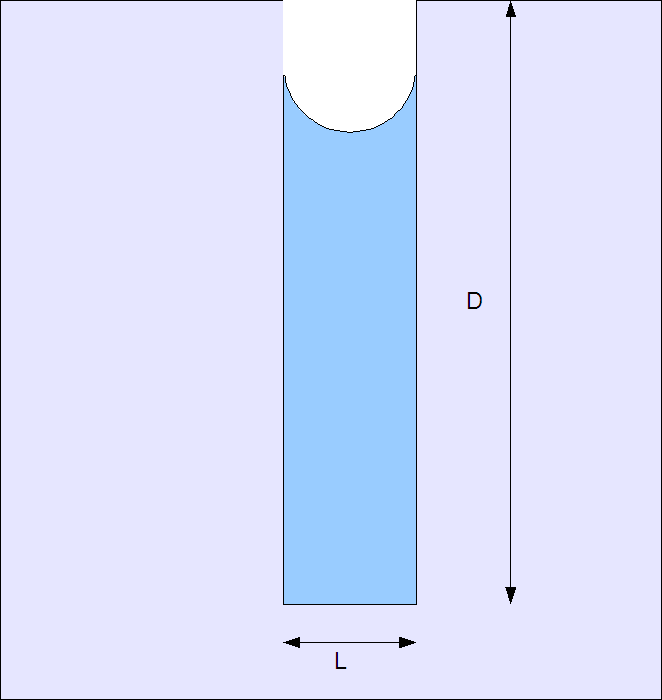}
 \caption{Sketch of a cross section of a capillary groove of width $L$ and depth $D\gg L$ and
three different configurations of the meniscus. On the far left is shown a near empty capillary corresponding to $\mu<\mu_{\rm cc}$ with two small corner menisci
near the capillary bottom. On the far right is a near filled capillary corresponding to $\mu<\mu_{\rm cc}$ in which a single meniscus is close to the groove
opening. The middle configuration shows an empty capillary corresponding to $\mu\approx\mu_{\rm cc}$ but with much more prominent corner menisci which may
correspond to a stable or a metastable configuration depending on the order of the condensation transition.}\label{scheme}
\end{figure*}

It is well known that confining a fluid can dramatically alter its properties and induce new examples of phase transitions and critical phenomena. Important
examples of this include wetting, pre-wetting and layering at planar walls \cite{sullivan, dietrich, schick, bonn}, capillary condensation, critical point shifts
and interfacial delocalization  in parallel plate geometries \cite{evans1, evans2, fisher, parry_cwalls} and filling transitions in wedges \cite{rejmer, wood1,
binder03, our_prl, our_wedge}. A simple extension of these idealized geometries, which has received a great deal of recent attention, is a groove or capped
capillary formed by scoring a narrow deep channel or array of channels into a solid surface (see Fig. 1). Capping a capillary strongly influences the nature of
condensation and evaporation, compared to that occurring in an infinite open capillary-slit, because of the presence of a meniscus, which must unbind from the
groove bottom (top) at condensation (evaporation) \cite{darbellay, evans_cc, tasin, mistura, hofmann, schoen, roth, mal_groove, parry_groove, ser, mal_13}. The
purpose of the present paper is to investigate the nature of this meniscus unbinding in a model microscopic density functional theory (DFT) incorporating
realistic long-ranged intermolecular forces. In doing so we wish to study a number of predictions for the possible asymmetry between the condensation and
evaporation branches of the adsorption isotherm occurring both above and below the wetting temperature \cite{parry_groove}. In particular, we report for the
first time microscopic studies of evaporation transitions in capillary grooves with long-range intermolecular forces and show a precise connection, or
covariance, with complete wetting at planar walls.

Consider a simple fluid which, in the bulk, shows coexistence between liquid and gas phases, with densities $\rho_l$ and $\rho_g$ respectively, along a saturation chemical potential curve $\mu_{\rm sat}(T)$ which terminates at
a bulk critical temperature $T_c$. Now suppose the fluid is confined between two parallel plates, of infinite area, separated by a distance $L$. Such confinement shifts the location of the coexistence, which now occurs between
capillary gas and capillary liquid phases along a capillary condensation curve $\mu_{\rm cc}(L)$ and which ends at a capillary critical temperature \cite{fisher, evans1}. In the limit of macroscopically wide capillaries and
small undersaturation, $\mu_{\rm cc}(L)$ satisfies the Kelvin equation \cite{thomson}:
 \bb
 \mu_{\rm cc}(L)=\mu_{\rm sat}-\frac{2\gamma\cos\theta}{L\Delta\rho}+\cdots\,,\label{kelvin}
 \ee
where $\Delta\rho=\rho_l-\rho_g$, $\gamma$ is the liquid-gas interfacial tension and $\theta$ is the contact angle. For temperatures below the wetting temperature $T_w$, corresponding to $\theta>0$, the Kelvin equation is
known to be accurate even for microscopically  narrow slits which are only several molecular diameters wide \cite{evans2}. However, for $T>T_w$, when  $\theta=0$ and thick wetting films form at the walls, it is necessary to
also allow for Derjaguin's correction which  reduces the effective slit width \cite{evans1, evans2, derj} (see Appendix). In this case the Derjaguin corrected Kelvin equation is known to be accurate down to slit widths of a
few hundred molecular diameters \cite{evans2}. In most DFT studies of capillary condensation one supposes translational invariance along the walls so that the density profile depends only on one Cartesian coordinate. In this
case capillary condensation is certainly a first-order transition, at which the density profile jumps, at $\mu_{cc}$, from a low coverage to a high coverage state. Thus at mean-field level, below the capillary critical
temperature $T_c(L)$, adsorption isotherms exhibit a van der Waals loop, characterised by metastable extensions and spinodals. The Kelvin equation is usually derived by equating the grand potentials of coexisting capillary
liquid and capillary gas phases each of which have distinct bulk and surface free-energy contributions \cite{evans2}. Alternatively, it can understood geometrically as the value of the chemical potential where a circular
meniscus of Laplace radius $R=\gamma/(\mu_{\rm sat}-\mu_{\rm cc})\Delta\rho$ meets the walls at the contact angle $\theta$, thus establishing a stable separation of the coexisting capillary phases.

This simple picture of capillary condensation is enriched considerably when one end is capped to form a deep capillary groove, as shown schematically in
Fig.~\ref{scheme}. Hereafter we suppose that the side walls and cap are made from the same material so there is only one macroscopic contact angle $\theta$. In
such a macroscopically deep groove condensation must still occur at $\mu_{\rm cc}(L)$ since the contribution to the free energy from the capped and open ends do
not scale with the depth $D$. However it is now necessary to distinguish between condensation and evaporation since these two transitions are distinct.
Condensation involves the unbinding of the meniscus from the bottom of the groove, which is filled mostly with capillary gas along the adsorption branch, as the
chemical potential is increased to $\mu_{\rm cc}$. Evaporation on the other hand involves the unbinding of the meniscus from the top of the groove, which is
filled mostly with capillary liquid along the desorption branch, as the chemical potential is decreased to $\mu_{\rm cc}$. Recent theoretical studies
\cite{evans_cc, schoen, roth} have predicted that for $T>T_w$ when the walls  are completely wet by liquid the condensation involves the continuous unbinding of
the meniscus. Thus, if $\ell_C$ denotes the equilibrium height of the meniscus above the groove bottom, we can characterise the divergence on approaching
condensation by
 \bb
 \ell_C\sim(\mu_{\rm cc}-\mu)^{-\beta_C}\,,\label{ell_cond}
 \ee
which is valid for deep grooves $D\gg\ell_C\gg L$. Similarly, for $T>T_w$ the evaporation transition is also predicted to be continuous \cite{parry_groove} in which case the distance of the meniscus from the groove top
diverges as
 \bb
 \ell_E\equiv(D-\ell_C)\sim(\mu-\mu_{\rm cc})^{-\beta_E}\,,\label{ell_ev}
 \ee
which is again valid for $D\gg\ell_E\gg L$. Strictly speaking these power laws only apply to macroscopically deep grooves but as we shall see they also describe
the behaviour of the meniscus in grooves of finite depth. The values of the exponents $\beta_C$ and $\beta_E$ (which are referred to as $\beta_A$ and $\beta_D$
in Ref. \cite{evans_cc}) are, in general,  distinct with both depending sensitively on the range of intermolecular forces. In particular, for dispersion-like
forces, effective Hamiltonian considerations lead to the mean-field predictions $\beta_C=1/4$ and$\beta_E=1/3$  which are almost entirely insensitive to
interfacial fluctuation effects (see later) \cite{evans_cc}. Thus even when the condensation and evaporation are both continuous there is still an asymmetry
between them with the condensation being a slightly sharper transition. Note that the value of the evaporation exponent $\beta_E$ is the same as that one
describing the growth of the complete wetting films at planar walls in systems with dispersion forces. In fact, this is not coincidental and points to a much
deeper connection between groove evaporation and planar complete wetting which we shall discuss at length in this paper.

The situation for walls which are partially wet by liquid is however quite different \cite{parry_groove, mal_groove, ser}. Below a temperature $T^*(L)\approx
T_w$ it has been predicted that the condensation transition becomes first-order due to the preferential adsorption of fluid, or corner menisci, at the bottom of
the groove (see the left panel in Fig.~\ref{scheme}).  In this case at $\mu_{\rm cc}$ a low coverage configuration coexists with one in which the groove is
completely filled with liquid (see the right panel in Fig.~\ref{scheme}) similar to the standard interpretation of condensation. Note that the macroscopic size
and shape of such corner menisci are determined uniquely by the simple geometrical requirement that they have a Laplace radius $R=\gamma/(\mu_{\rm
sat}-\mu)\Delta\rho$ and meet the side and bottom walls at the correct contact angle $\theta$. Such low coverage configurations persist into a metastable regime
$\mu>\mu_{\rm cc}$ which terminates at a spinodal $\mu_{\rm sp}$ where the menisci meet, determined by the complementary Kelvin equation \cite{parry_groove}
 \bb
 \mu_{\rm sp}(L)=\mu_{\rm cc}(L)+\frac{2\gamma\sin\theta}{L\Delta\rho}+\cdots\,.\label{ckelvin}
 \ee

The phenomena that we focus on in this paper are the condensation and evaporation transitions and in particular their order above and below the wetting
temperature $T_w$. To fully understand them we must also consider them in the context of other phase transitions both at mean-field level and beyond. As
mentioned earlier, the groove is itself a generalization of other geometries which exhibit known and well-studied phase transitions. Suppose for example that the
separated side walls exhibit (in the limit $L\to \infty$) a first-order wetting transition at temperature $T_w$ and chemical potential $\mu=\mu_{\rm sat}$.
Associated with this is a pre-wetting line which extends above $T_w$ and to $\mu<\mu_{\rm sat}$ which is the locus of coexistence between two distinct phases
with thin and thicker wetting layers. In a capillary of finite width $L$, the wetting transition is of course suppressed but a finite-size shifted capillary
pre-wetting line will still, in general, exist. In this case one should distinguish between capillary-gas phases which have either thin or thick wetting films at
the side walls. However for narrow capillaries, which will be exclusively considered in this paper, the capillary pre-wetting occurs in a metastable region since
it is preceded by capillary condensation. Similarly, the bottom of a capped capillary comprises two right-angle corners each of which, if separated by an
infinite distance, would, at bulk two phase coexistence, exhibit a filling transition at a temperature $T_f<T_w$ (when the contact angle $\theta=\pi/4$). In a
capillary of finite width $L$ the filling transition is again suppressed since the corner menisci can not become macroscopically large. But, if the filling
transition is first-order one may suspect that a pre-filling line, corresponding to transitions from thin to thick corner menisci states, exists at either corner
of a finite width capillary. However there are two reasons why such transitions are irrelevant for our discussion of condensation and evaporation. First,
analogous to pre-wetting at the side walls, for narrow grooves, the pre-filling line exists in a metastable region of the phase diagram since it is preceded by
capillary condensation. The second reason is that, strictly speaking, the pre-filling transition is an artefact of the mean-field approximation in model density
functional theories. In reality, any pre-filling-like jump in the adsorption near the corners is rounded owing to the pseudo one-dimensional nature of the
transition. Exactly the same reasoning applies to any pre-wetting like transition associated with the bottom wall which must also be rounded beyond mean-field.
We will return to this last point at the end of our paper.  In summary, pre-wetting and pre-filling associated with the bottom wall are irrelevant artefacts of
mean-field treatments, while pre-wetting at the side walls is only of interest for rather wide capillaries.

In this paper we use a highly accurate fundamental measure DFT to examine different properties of condensation and evaporation in a narrow groove geometry for
systems with long-ranged intermolecular forces. We aim to answer the following questions; Firstly regarding continuous capillary condensation above $T_w$, is the
effective Hamiltonian prediction that the $\beta_C=1/4$ observable in our studies of grooves of large but finite depth? Secondly does the transition become first
order below the wetting temperature and if so how accurate is the complementary Kelvin equation? Similarly for evaporation we wish to test the prediction that
above $T_w$ the exponent $\beta_E=1/3$ and to determine the order of the transition below $T_w$. In doing this we will show that due to the presence of
long-ranged forces, for capillary evaporation there is a hidden connection or covariance with the very well understood phenomena of complete wetting at a planar
wall i.e. the adsorptions characterising these different phase transitions in two distinct geometries are precisely related to each other.


The rest of our paper is organised as follows. In section II we describe our DFT, our choice of intermolecular forces and the substrate geometry in more detail.
In Section III we present our DFT results for condensation and evaporation below (IIIa) and above (IIIb) the wetting temperature and determine numerically the
critical exponents $\beta_C$ and $\beta_E$. We also show that above the wetting temperature, where corner menisci are not of crucial importance, the values of
the critical exponents can be obtained analytically from a simple slab or shark-kink model the details of which are provided in an Appendix. Finally, we discuss
in detail the covariance between groove evaporation and complete wetting on a planar wall and finish with a summary of our results and a discussion of the nature
of the capillary wetting transition beyond mean-field.

\section{Density Functional Theory}

Within DFT \cite{evans79}, the equilibrium density profile is found by minimising the grand potential functional
 \bb
 \Omega[\rho]={\cal F}[\rho]+\int\dd\rr\rhor[V(\rr)-\mu]\,,\label{om}
 \ee
where $\mu$ is the chemical potential and $V(\rr)$ is the external potential. It is convenient to divide the intrinsic free energy functional ${\cal F}[\rho]$
into an exact ideal gas contribution and an excess part:
  \bb
  {\cal F}[\rho]=\int\dr\rho(\rr)\left[\ln(\rhor\Lambda^3)-1\right]+{\cal F}_{\rm ex}[\rho]\,,
  \ee
where $\Lambda$ is the thermal de Broglie wavelength, which we set to unity. To continue we follow the traditional van der Waals or perturbative approach and model the excess term as a sum of hard-sphere and attractive
contributions where the latter is treated in a simple mean-field fashion. Thus we write
  \bb
  {\cal F}_{\rm ex}[\rho]={\cal F}_{\rm hs}[\rho]+\frac{1}{2}\int\int\dd\rr\dd\rr'\rhor\rho(\rr')u_{\rm a}(|\rr-\rr'|)\,, \label{f}
  \ee
where  $u_{\rm a}(r)$ is the attractive part of the fluid-fluid interaction potential.

Minimisation of (\ref{om}) leads to an Euler-Lagrange equation:
 \bb
 V(\rr)+\frac{\delta{\cal F}_{\rm hs}[\rho]}{\delta\rho(\rr)}+\int\dd\rr'\rho(\rr')u_{\rm a}(|\rr-\rr'|)=\mu\,.\label{el}
 \ee

In our model, the fluid atoms are assumed to interact with each other via a truncated (but non-shifted) Lennard-Jones-like potential
 \bb
 u_{\rm a}(r)=\left\{\begin{array}{cc}
 0\,;&r<\sigma\,,\\
-4\varepsilon\left(\frac{\sigma}{r}\right)^6\,;& \sigma<r<r_c\,,\\
0\,;&r>r_c\,.
\end{array}\right.\label{ua}
 \ee
which is cut-off at $r_c=2.5\,\sigma$, where $\sigma$ is the hard-sphere diameter. Hereafter, we will use the parameters $\sigma$ and $\varepsilon$ as the length and energy units.

The hard-sphere part of the excess free energy is approximated by the fundamental measure theory (FMT) functional \cite{ros},
 \bb
{\cal F}_{\rm hs}[\rho]=\frac{1}{\beta}\int\dd\rr\,\Phi(\{n_\alpha\})\,,\label{fmt}
 \ee
where  $\beta=1/k_BT$ is the inverse temperature. There exist various recipes for constructing FMT functionals in terms of the weighted densities $n(\alpha)$,
all of which are known to accurately model the short-rage correlations. Here we follow the modified version of the Rosenfeld original functional proposed in
Ref.~\cite{mal_dft}, which is known to satisfy exact statistical mechanical sum rules and thermodynamic conditions at planar walls and corners \cite{our_prl,
our_wedge}.

To construct the external potential $V(\rr)$ we consider a semi-infinite solid slab of uniformy density $\rho_w$, into which is cut an infinitely long narrow
groove of width $L$ and depth $D$ as shown in Fig.1. The wall atoms interact with the fluid particles via the attractive part of Lennard-Jones potential
 \bb
 \phi(r)=-4\varepsilon_w\left(\frac{\sigma}{r}\right)^{6}\,,\label{ext_field}
 \ee
so that total external potential experienced by the fluid atoms inside the groove is
 \begin{widetext}
 \bb
 V(x,z; L, D)=\left\{\begin{array}{ll} \infty,&x<\sigma\;{\rm or}\;x>L-\sigma\;{\rm or}\;z<\sigma,\\
\rho_w\int\phi(|\rr-\rr'|)\dd r'& {\rm elsewhere\,,}\end{array}\right.\label{V}
 \ee
\end{widetext}
where the integral is over the solid volume and we have incorporated a hard-wall repulsion set by the atomic diameter $\sigma$ which prevents $V(x,z)$ from
diverging. Thus the external potential is translationally invariant along the $y$-axis running parallel to the groove. The integrals can be done analytically and
the potential can be written as a sum of contributions from the bottom wall and vertical walls on the LHS and RHS respectively:
 \bb
  V(x,z; L,D)=V^{(1)}(z)+V^{(2)}(x,z; D)+V^{(2)}(L-x,z; D)\,,\label{Vw}
 \ee
The potential due to the bottom wall is particularly simple and decays as a pure power law,
 \bb
 V^{(1)}(z)=\frac{2\alpha_w}{z^3}\label{Vw1}\,,
 \ee
where we have introduced the pre-factor
 \bb
 \alpha_w=-\frac{1}{3}\pi\varepsilon_w\rho_w\sigma^6\,.
 \ee
The contribution due to each vertical wall is more complicated and is most conveniently written as\cite{mal_grooves}
 \bb
 V^{(2)}(x,z; D)=\alpha_w\left(\psi(x,z)+\psi(x,D-z)\right)\,, \label{Vw2}
 \ee
where
 \bb
\psi(x,z)=\frac{2z^4+x^2z^2+2x^4}{2x^3z^3\sqrt{x^2+z^2}}-\frac{1}{z^3}\,. \label{psi}
 \ee
It is straightforward to show that in the limits $L\to\infty$ and $D\to\infty$ the potential $V(x,z)$ reduces to that corresponding to a rectangular corner
\cite{our_wedge}.

Using the external potential $V(x,z)$ we numerically solve the Euler-Lagrange equation (\ref{el}) for the equilibrium profile $\rho(x,z)$ on a two-dimensional Cartesian grid with a spacing $0.05\,\sigma$.
In our numerical studies we set the wall strength $\varepsilon_w=1.2\,\varepsilon$.  In this case it is known that, for a planar wall-gas interface at bulk
coexistence $\mu=\mu_{\rm sat}$, there is a strongly first-order wetting transition at a temperature satisfying $k_BT_w=1.18\,\varepsilon$ which is far below the
bulk critical temperature  (occurring at $k_BT_c=1.41\,\varepsilon$) \cite{our_wedge}. Similarly for a single right-angle corner there exists a strongly
first-order filling transition occurring also at bulk coexistence but at a lower temperature satisfying $k_BT_f=1.08\,\varepsilon$. Before we considered the
capped groove we first studied an infinite open slit for which the external potential $V(x)=2\alpha_w( x^{-3}+(L-x)^{-3})$. For this system we determined the
equilibrium grand potentials and thus computed the capillary coexistence curve $\mu_{\rm cc}$ as a function of $T$ for representative widths $L=7\sigma$ and
$L=12\sigma$. We then cap this geometry leaving an open end in order to study the nature of the condensation and evaporation transitions. For this we chose a
groove with depth $D=50\sigma$ which is sufficiently deep to observe the meniscus unbinding and accurately determine critical exponents for these transitions. To
model the boundary with the bulk reservoir at the top of the groove we use the simple boundary condition $\rho(x,D)=\rho_b\exp[-\beta V(x,D; L,D)]$ where
$\rho_b$ is the bulk vapour density. This precludes us studying wetting films on top of the sculpted surface but does not influence the condensation and
evaporation occurring within a deep groove for chemical potentials close to $\mu_{\rm cc}$. Having determined the density profile $\rho(x,z)$ we construct the
total adsorption
 \bb
 \Gamma_C=\int\int\dd x \dd z (\rho(x,z)-\rho_b)\,,\label{ads}
 \ee
which directly measures the height of the meniscus from the bottom and is thus an appropriate order parameter for condensation in a groove. For evaporation, it
is more suitable to consider an excess over the density of the (metastable) bulk liquid density $\rho_l^+$:
  \bb
 \Gamma_E=\int\int\dd x \dd z |\rho(x,z)-\rho_l^+|\,,\label{des}
 \ee
which is proportional to the height of the meniscus from the top. We have determined adsorption isotherms of $\Gamma_C$ and $\Gamma_E$ vs. $\mu$, for a variety
of temperatures above and below $T_w$ which directly reflect the order of the condensation and evaporation transitions. For condensation we start from a low
coverage configuration corresponding to $\mu\ll\mu_{\rm cc}$ and increase the chemical potential until either a spinodal is reached or the groove continuously
fills. For evaporation we consider the reverse scenario and start from a high coverage at a chemical potential $\mu>\mu_{\rm cc}$ (but still below the saturation
value $\mu_{\rm sat}$) and approach $\mu_{\rm cc}$ from above.

\section{Results}

\subsection{Condensation and Evaporation below the Wetting Temperature}

\begin{figure}[ht]
\includegraphics[width=8cm]{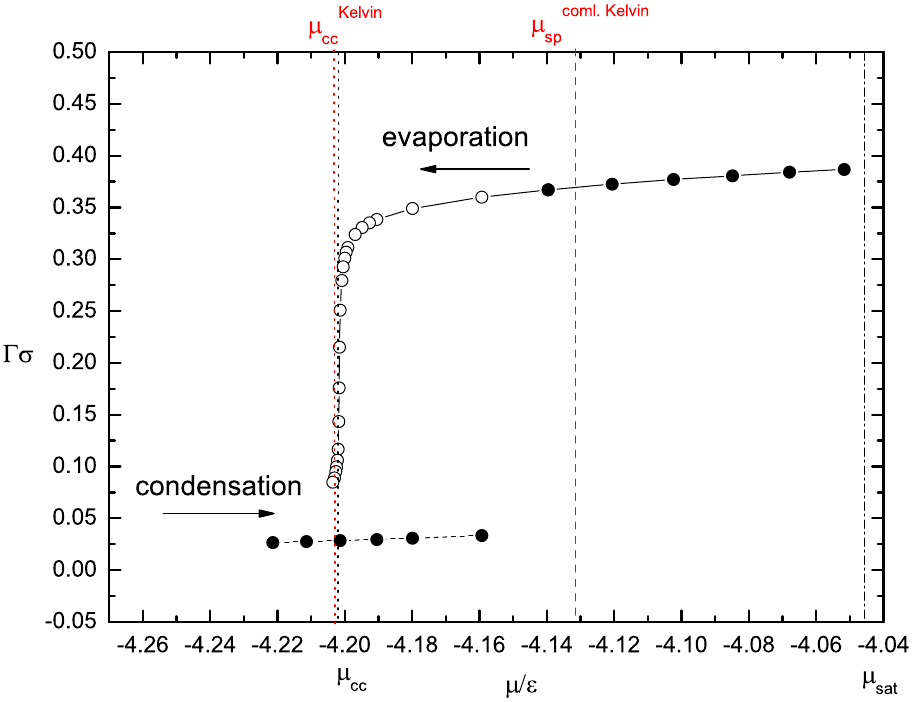}
\caption{Adsorption isotherms (full symbols) found by increasing $\mu$ and desorption isotherms (empty symbols) found by decreasing $\mu$ in a capillary groove
of width $L=7\sigma$ and depth $D=50\sigma$, at a temperature $T/T_w=0.97$. The black vertical dotted line denotes the value of the chemical potential
corresponding to the capillary condensation determined independently from a 1D DFT for an open slit of width $L=7\sigma$; the red vertical dotted lines denotes
the chemical potential corresponding to the capillary condensation determined from the Kelvin equation, Eq.~(\ref{kelvin}).  The red dashed line denotes the
position of the spinodal as predicted by the complementary Kelvin equation, Eq.~(\ref{ckelvin}). The vertical dotted-dashed line denotes the saturated chemical
potential.}\label{cond_ev}
\end{figure}

\begin{figure}[ht]
\includegraphics[width=6cm]{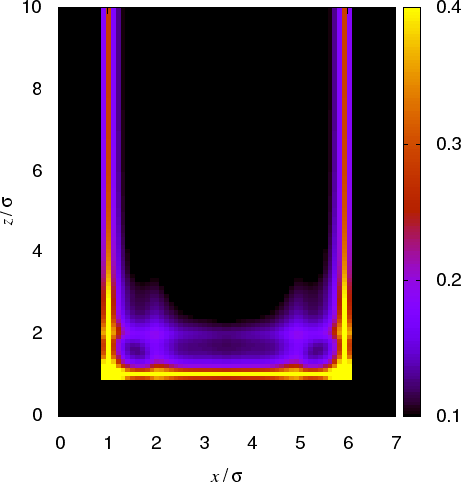}
\caption{Density profile corresponding to a low-adsorption state in a groove of $L=7\sigma$ and $D=50\sigma$ (displayed only to $z=10\sigma$) at $T/T_w=0.97$ at
capillary coexistence $\mu=\mu_{\rm cc}$.}
\label{prof_T114}
\end{figure}

\begin{figure*}[ht]
\includegraphics[width=2cm]{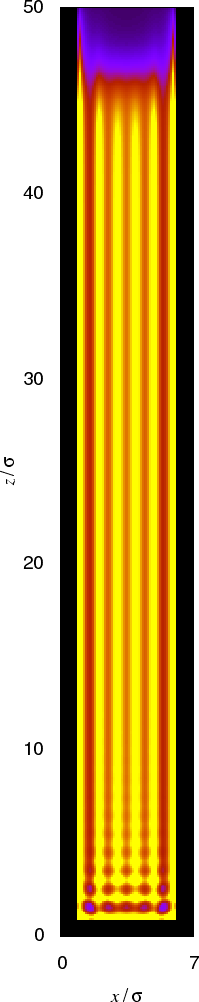} \hspace{3cm}\includegraphics[width=2cm]{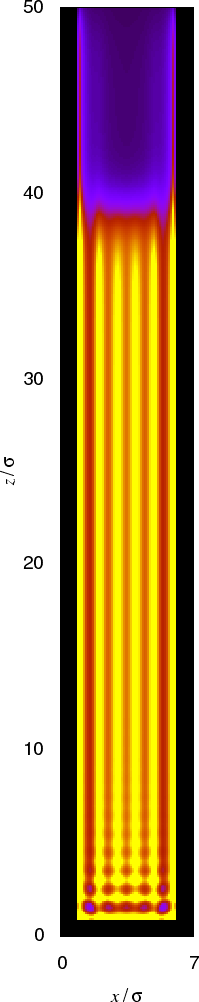}\hspace{3cm} \includegraphics[width=2.44cm]{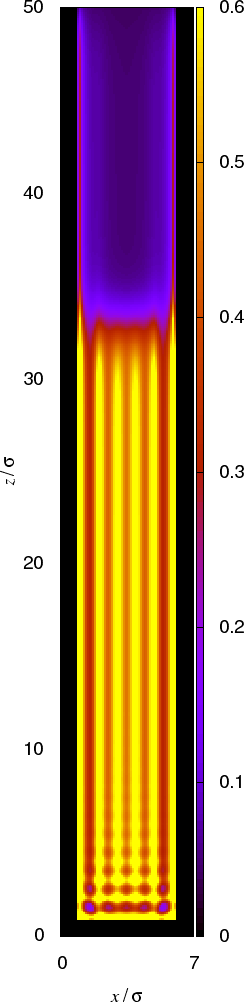}
\caption{Density profiles corresponding to the evaporation branch for a capillary groove of width $L=7\sigma$ and depth $D=50\sigma$, at a temperature
$T/T_w=0.97$. From left to right the undersaturation $\delta\mu=\mu-\mu_{\rm sat}$ is: $10^{-3}$, $10^{-4}$, and $5\cdot10^{-5}$ in units of $\varepsilon$. }
\label{profs_ev_T114}
\end{figure*}

\begin{figure}[ht]
\includegraphics[width=8cm]{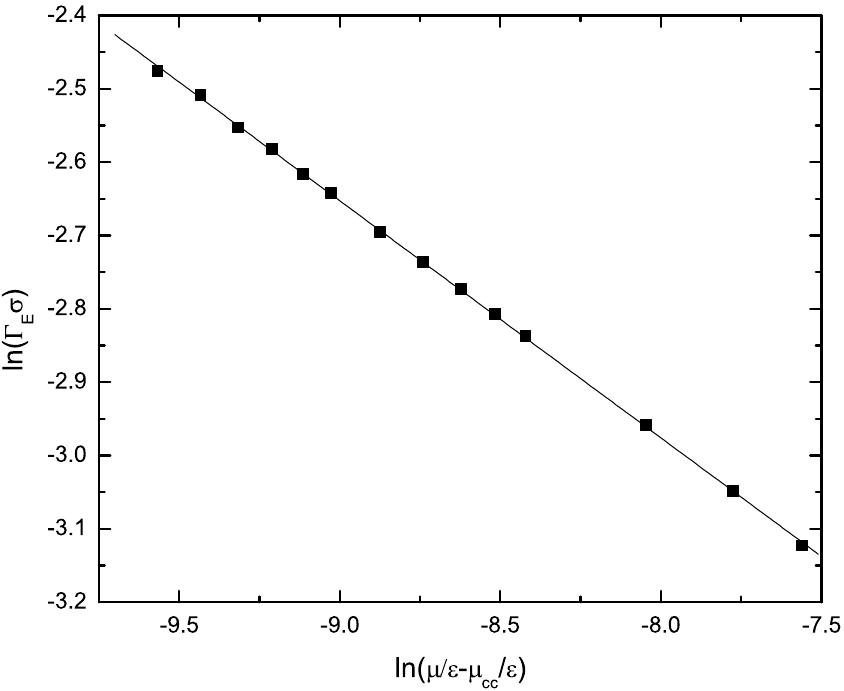}
\caption{A log-log plot of the adsorption $\Gamma_E$ for the evaporation branch, $\mu\to\mu_{\rm cc}^+$ for a groove of width $L=7\sigma$ and depth $D=50\sigma$
at sub-wetting temperature $T/T_w=0.97$. The straight line has a slope $-1/3$.} \label{ev_crit}
\end{figure}

In Fig.~\ref{cond_ev} we show a representative adsorption and desorption isotherms in a groove of width $L=7$ and depth $D=50$ at a temperature $T/T_w=0.97$. Consider first the condensation branch which follows the adsorption
as the chemical potential is increased from an initial low value. It can be seen that the coverage remains small as one approaches the chemical potential $\mu_{\rm cc}$ at which, in an infinitely deep groove, there is
coexistence with a high density configuration. The density profile at $\mu_{\rm cc}$ is shown in Fig.~\ref{prof_T114} and illustrates some preferential adsorption near the corners. There is however no macroscopic meniscus. We
note that the numerically determined value for $\mu_{\rm cc}$ is extremely close to that predicted by the Kelvin equation, which is to be anticipated since $T<T_w$. For $\mu>\mu_{\rm cc}$ the low density configuration is
metastable with respect to a high coverage state and persists up to a spinodal value $\mu_{\rm sp}$. The numerically obtained value $\mu_{\rm sp}=-4.15\varepsilon$ is also close, within $0.5\%$, to that predicted by the
complementary Kelvin equation (\ref{ckelvin}), although the agreement is not as good as for that between $\mu_{\rm cc}$ and the Kelvin equation. The accuracy of the complementary Kelvin equation  is remarkable in view of the
fact that it was derived using entirely macroscopic arguments based on the merging of corner menisci even though these are not present at these microscopic scales.

Consider next moving along the desorption line starting from a high-density state we observe a continuous and dramatic decrease in $\Gamma$. Representative
density profiles for different $\mu$ are shown in Fig.~\ref{profs_ev_T114} and show a meniscus whose distance from the open end increases continuously as
$\mu\to\mu_{\rm cc}$ . A log-log plot of $\Gamma_E$ vs $\mu-\mu_{\rm cc}$ is shown in Fig.~\ref{ev_crit} and is consistent with the result $\beta_E=1/3$.  We
will return to this in the next section.

These numerical results show that within our model of a macroscopically deep groove, below the wetting temperature, the condensation is first-order while the
evaporation is continuous. Of course, in a groove of finite depth the evaporation transition is a subject to finite size effects which limit the distance of the
meniscus from the open end. In this case, at mean-field level, the evaporation branch must also reach a spinodal $\mu_{\rm sp}^{\rm evap}$ which lies slightly
below $\mu_{\rm cc}$ (see Fig. 2). However, as $D$ becomes larger, $\mu_{\rm sp}^{\rm evap}$ tend to $\mu_{\rm cc}$. It is in this sense that the evaporation
transition is ultimately continuous in a macroscopically deep groove.

In contrast, the spinodal point $\mu_{\rm sp}^{\rm cond}$, corresponding to the condensation branch, remains distinct from $\mu_{\rm cc}$ as $D$ is increased
indicating that the transition is first-order. In fact for condensation one needs to consider the opposite limit and ask what happens to the transition as the
groove depth is decreased.  We have checked numerically that, for a variety of temperatures below $T_w$, the condensation transition remains first-order until
the depth $D$ becomes microscopically small, comparable to the width $L$.

Beyond mean-field level the condensation/evaporation transition is rounded for all finite $D$ since the groove geometry is pseudo-one-dimensional. However
standard finite-size scaling arguments imply that the width of the rounding $\Delta\mu_{\rm round}/\mu_{\rm cc}\approx \exp(-\gamma\beta LD)$ which is completely
negligible once the groove depth and width are greater than the bulk correlation length.

\subsection{Condensation and Evaporation above the Wetting Temperature}

\subsubsection{Numerical DFT results}

\begin{figure}[ht]
\includegraphics[width=8cm]{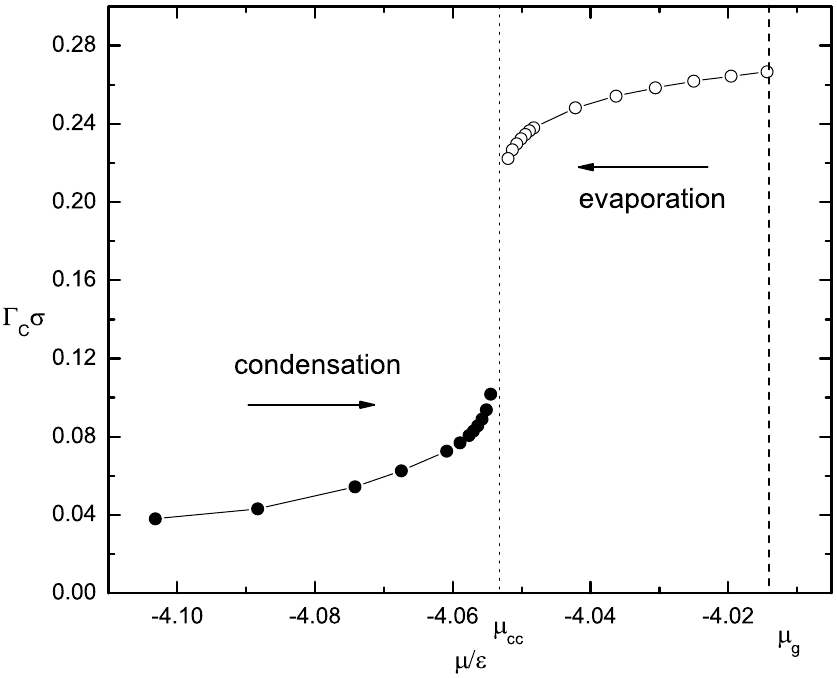}
\caption{Adsorption isotherm in a capillary groove of width $L=12\sigma$ and depth $D=50\sigma$, at a temperature $T/T_w=1.1$. The vertical dotted line denotes
the location of the chemical potential $\mu_{\rm cc}$   corresponding to the capillary condensation determined independently from the 1D DFT for an open slit of
width $L=12\sigma$; the vertical dashed line denotes saturated vapour density at the same temperature.}\label{T13}
\end{figure}

\begin{figure*}[ht]
\includegraphics[width=2cm]{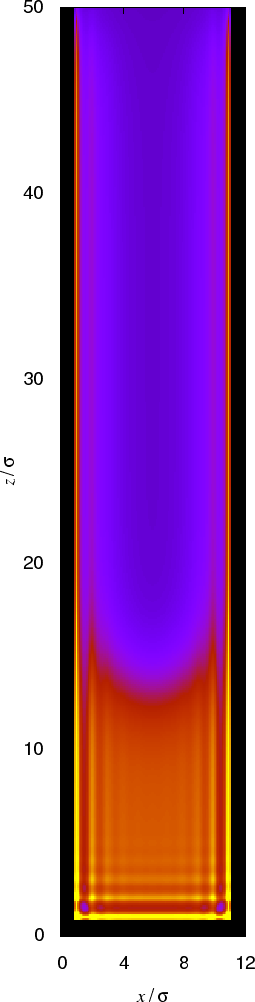} \hspace{2cm}\includegraphics[width=2cm]{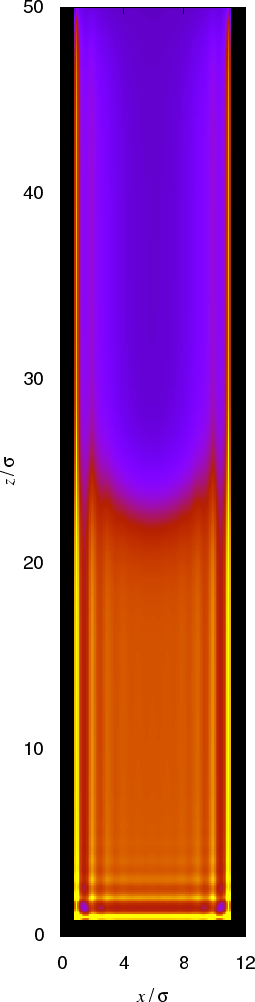}\hspace{2cm} \includegraphics[width=2.35cm]{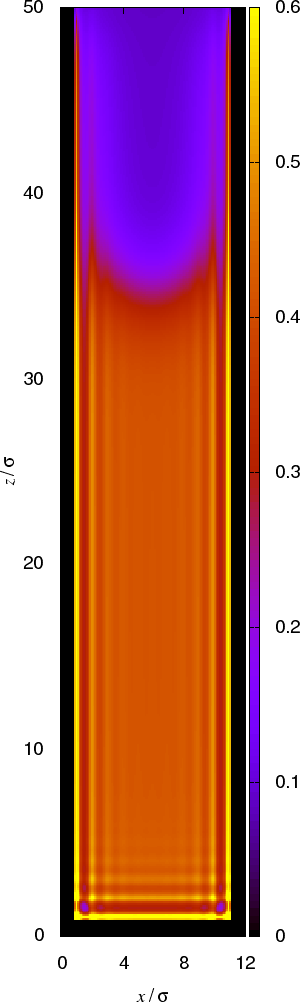}
\caption{From left to right density profiles for chemical potentials slightly below ($\mu/\mu_{\rm cc}=0.98$, close to ($\mu/\mu_{\rm cc}\approx1$) and slightly
above ($\mu/\mu_{\rm cc}=1.02$) that of capillary condensation for a capillary groove of width $L=12\sigma$ and depth $D=50\sigma$, at a temperature
$T/T_w=1.1$.} \label{profs_T13}
\end{figure*}

\begin{figure*}[ht]
\includegraphics[width=8.85cm]{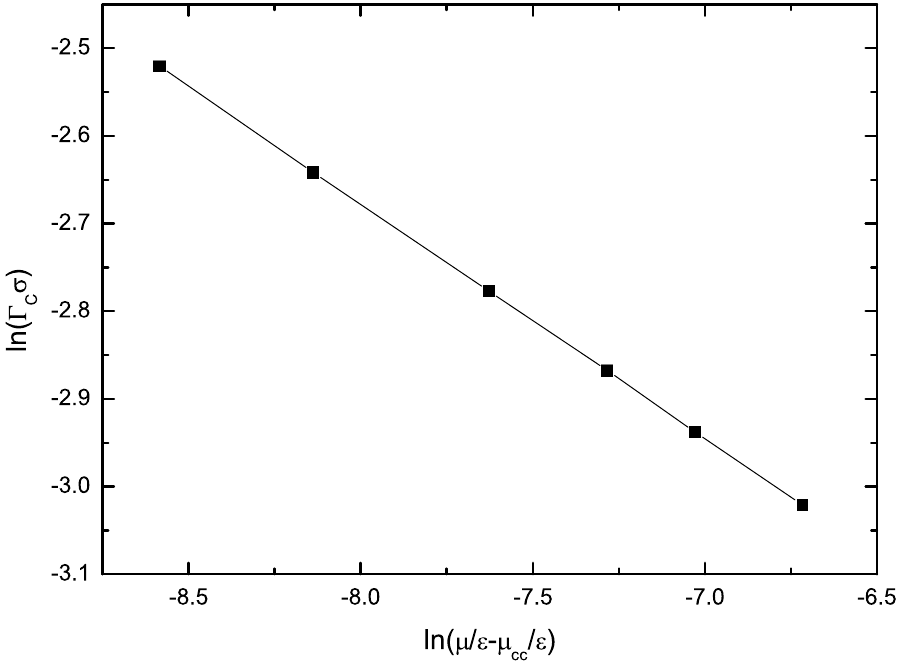} \includegraphics[width=8cm]{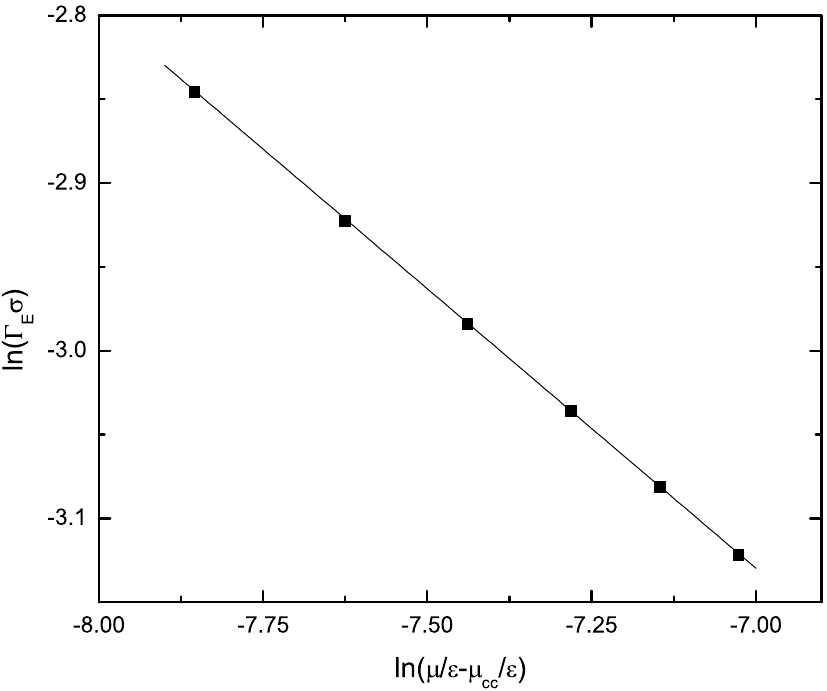}
\caption{Left: a log-log plots of the adsorption, $\Gamma_C$, for the condensation transition occurring  as $\mu\to\mu_{\rm cc}^-$. The straight line fit has
gradient $-0.2507$ (compare to the theoretical prediction $\beta_C=1/4$). Right:  a log-log plots of the complementary adsorption, $\Gamma_E$, for the
evaporation transition occurring  as $\mu\to\mu_{\rm cc}^+$. The straight line fit has gradient $-0.3332$ (compare to the theoretical prediction $\beta_E=1/3$).
These results pertain to a capillary groove of width $L=12\sigma$ and depth $D=50\sigma$, at a temperature $T/T_w=1.1$.} \label{crit_exp_T13}
\end{figure*}

In Fig.~\ref{T13} we show the adsorption and desorption isotherms obtained for $T/T_w=1.1$ in a groove of width $L=12\sigma$ and depth $D=50\sigma$. In this case
there is no hysteresis so that the condensation and evaporation branches are connected continuously. This finding is consistent with the prediction of the
complementary Kelvin equation since the contact angle $\theta=0$. In Fig.~\ref{profs_T13} we show density profiles obtained below, at, and slightly above
$\mu_{\rm cc}$, which illustrate the continuous movement of the meniscus from the near cap to groove top as the chemical potential is increased. The absence of
spinodals indicates that in the limit of macroscopic $D$ both condensation and evaporation transitions occur via  the continuous unbinding of the meniscus from
the bottom and top, respectively. However, while both transitions are continuous there remains a quantitative difference between them characterised by the
critical exponents $\beta_C$ and $\beta_E$. This is shown in Fig.~\ref{crit_exp_T13} where we present log-log plots of the adsorption and complementary
adsorption as $\mu_{\rm cc}$ is approached from below and above, respectively. For the evaporation branch, we find the same value $\beta_E=1/3$ obtained for the
case of $T<T_w$. For condensation on the other hand our results indicate that $\beta_C=1/4$.

\subsubsection{Slab model analysis}

Our numerical DFT results illustrate the asymmetry between condensation and evaporation. Below the wetting temperature condensation is first-order while the
evaporation is continuous. This qualitative difference can understood due to the behaviour of the corner menisci and is accurately quantified, even for narrow
slits, by the complementary Kelvin equation. Above the wetting temperature on the other hand the difference between condensation and evaporation is more subtle
since both transitions are continuous. However an asymmetry still persists through the distinction between the adsorption critical exponents $\beta_C$ and
$\beta_E$. In this section we present details of a sharp-kink or slab model calculation, valid above the wetting temperature, which predicts, analytically, the
values of the critical exponents. This is similar to the original effective Hamiltonian analysis of continuous condensation and evaporation presented in
\cite{evans_cc} but improves on by accounting for thick complete wetting films which will allow us to derive a formula for $\mu_{\rm cc}$  consistent with the
Derjaguin corrected Kelvin equation.

In sharp-kink approximation we assume that the full two dimensional density profile $\rho(x,z)$ simply arises from a) a flat meniscus constrained to be at height
$\ell$ above the groove bottom, b) wetting films of thickness $\ell_\pi$ at each wall lying above the meniscus. Hence vapour at pressure $p$ occupies a volume
$V_g$ while a metastable bulk liquid at pressure $p_l^+$ is adsorbed at the capillary walls and below the meniscus (see Fig. 9). Such a parameterization neglects
the shape of the meniscus which is approximately circular. However, allowing for a fixed circular shape adds only a constant term to the free-energy and thus
does not influence in any way the equilibrium values of the meniscus height and wetting film thickness. More importantly this parameterization does not allow for
corner menisci and is therefore not appropriate for modelling condensation occurring below the wetting temperature.

For a macroscopically deep groove, the equilibrium value of the wetting film thickness at the side walls is independent of the meniscus height and is the same as
that for an infinite open slit, given by
 \bb
 \ell_\pi=\left(\frac{2|\alpha_w|}{\mu_{\rm sat}-\mu}\right)^\frac{1}{3}+{\cal{O}}(L^{-\frac{5}{3}})\,.\label{lpi}
 \ee
The first term here is the equilibrium film thickness of a complete wetting film at a single planar wall while the ${\cal{O}}(L^{-\frac{5}{3}})$ correction
arises from the interaction between the wetting films on the opposite walls which can be safely ignored. Using this value for $\ell_\pi$ we now substitute the
trial density profile into the functional $\Omega[\rho]$ and obtain a grand potential $\Omega(\ell)$ which is a function of the meniscus height. Per unit length
$L_\parallel$ of the groove, the constrained, excess, contribution to the grand potential density $\omega(\ell)=\Omega(\ell)/L_\parallel$ is given by

\begin{figure}[ht]
\includegraphics[width=5cm]{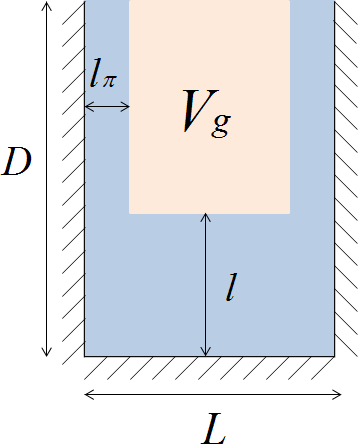}
\caption{Slab model parameterization of the density profile showing a meniscus of height $\ell$ in a capped capillary groove of width $L$ and height $D$
appropriate above the wetting temperature. $V_g$ denotes the volume filled by gas at pressure $p$. The side walls are coated with complete wetting films of
thickness $\ell_\pi$. }\label{groove}
\end{figure}

\begin{widetext}
\begin{eqnarray}
\omega^{\rm ex}(\ell)=(p_l^+-p)(L-2\ell_\pi)(D-\ell) +\gamma\left[2(D-\ell)+(L-2\ell_\pi)\right]-\frac{\Delta\rho}{L_\parallel}\int_{V_g}V(\rr)\dr\,.
\label{omex}
\end{eqnarray}
\end{widetext}
where we have defined the relevant excess contribution by subtracting off a constant term corresponding to the free-energy of a completely filled groove.
Minimization of $\omega^{\rm ex}$ determines the equilibrium meniscus height $\ell_C$ and $\ell_E$, details of which are presented in the Appendix.
In the limit $D\to\infty$ the minimization also determines the value of the chemical potential at which
condensation/evaporation occurs as
  \bb
  \mu_{\rm cc}(L)\approx\mu_{\rm sat}-\frac{2\gamma}{\Delta\rho(L-3\ell_\pi)}\,, \label{kelvin_slab}
  \ee
which is precisely the Derjaguin corrected Kelvin equation which allows for the shift in condensation due to thick complete wetting layers at the vertical walls.

For the condensation branch, occurring as $\mu$ approaches $\mu_{\rm cc}$ from below, we find that in the limit $D\to \infty$ of a macroscopically deep groove,
the equilibrium height of the meniscus above the groove bottom is \cite{mal_groove}
  \bb
\ell_C=\left(\frac{9|\alpha_w|L}{8(\mu_{\rm cc}-\mu)}\right)^\frac{1}{4}+\cdots \label{cond}
  \ee
where the ellipsis denote negligible non-diverging higher order terms. Similarly for the evaporation branch, occurring as $\mu$ approaches $\mu_{\rm cc}$ from
above, we find that in the limit of a macroscopically deep groove, the equilibrium height of the meniscus from the groove opening is
 \bb
 \ell_E=\left(\frac{2|\alpha_w|}{\mu-\mu_{\rm cc}}\right)^\frac{1}{3}+\cdots\,.\label{le}
 \ee
where we have again ignored  non-diverging terms.

At this point we make the following remarks:\\

i) The values of the exponents $\beta_C=1/4$ and $\beta_E=1/3$ are exactly the same as those predicted in the effective Hamiltonian study of \cite{evans_cc}
which adopted a slightly simpler parameterization of the density profile. Thus as expected the only influence of the thick wetting films at side walls is change
the location of the capillary condensation, $\mu_{cc}$, in keeping with Derjaguin corrected Kelvin equation. The values of these exponents, obtained in a slab
model for an infinitely deep groove, are in excellent agreement with our numerical DFT results obtained for the condensation and evaporation in a groove of
finite depth $D=50\sigma$.\\

ii) As discussed earlier the present slab model calculation cannot be used to study condensation occurring below $T_w$ since the parameterization of the profile
does not allow for corner menisci. However it can used to study evaporation for $T<T_w$ since in the limit of an infinitely deep groove, for all $\mu>\mu_{\rm
cc}$, the meniscus is always far from the groove bottom. Thus as for the case $T>T_w$ there is only a single meniscus. Because the the vertical walls are now
partially wet there is no need to allow for thick wetting films and one can set the parameter $\ell_\pi=0$ in the slab model analysis. In this case it is easy to
show that the result for the divergence of the meniscus depth remains $\ell_E=(2|\alpha_w|/(\mu-\mu_{\rm cc}))^{1/3}$. The only difference is that the location
of capillary condensation, $\mu_{\rm cc}$, is given by the by standard macroscopic Kelvin equation (\ref{kelvin}). Thus for our present system, with long-ranged
wall-fluid and short-ranged fluid-fluid interactions, the slab model predicts that evaporation remains continuous even below the wetting temperature $T_w$. This
is completely consistent with our numerical DFT results. We shall to return this point later where we shall try to generalise the criteria for the order of the
evaporation transition.

\subsection{Covariance between groove evaporation and complete wetting}

\begin{figure}[ht]
\includegraphics[width=8cm]{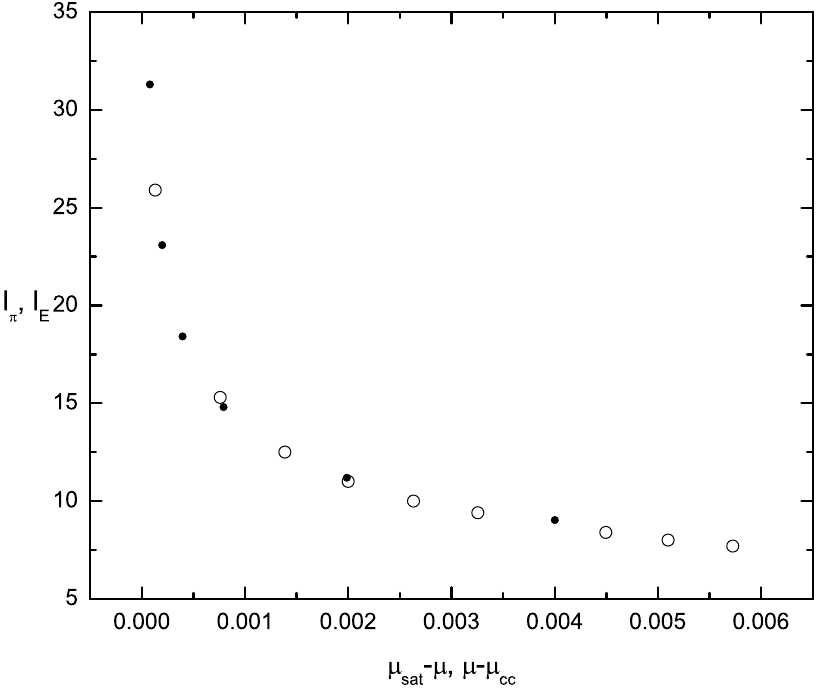}
\caption{Test of the covariance law showing comparison of the growth of the wetting layer thickness $\ell_\pi$ for complete wetting on a planar wall (open symbols) with the meniscus position $\ell_E$ for evaporation in a
capillary groove (filled symbols) of width $L=12\sigma$ and depth $D=50\sigma$ at temperature $T/T_w=1.1$. The results are expressed in the molecular units of $\sigma$ and $\varepsilon$.} \label{ev_cw}
\end{figure}

The slab model analysis points to a remarkably simple connection between groove evaporation, occurring for $T>T_w$, in systems with dispersion forces, and
complete wetting at a planar wall. Recall that the complete wetting transition occurs above the wetting temperature and refers to the divergence of the
equilibrium thickness, $\ell_\pi(\mu)$, of an adsorbed liquid film, at a planar wall-gas interface, as the chemical potential $\mu$ of the bulk gas , is
increased to saturation $\mu_{\rm sat}$. In general one writes this divergence as $\ell_\pi(\mu)\approx |\mu-\mu_{\rm sat}|^{-\beta_s^{\rm co}}$ where the
exponent $\beta_s^{\rm co}$ is determined by the range of the intermolecular forces and also possible interfacial fluctuation effects. For three dimensional
systems, fluctuations are negligible, and for dispersion (van der Waals) forces the exponent takes the mean-field value $\beta_s^{\rm co}=1/3$ first explained by
the Russian school of Derjaguin and Frumkin \cite{derj}. However, from the slab model results,
 \bb
 \ell_E(\mu)\approx\left(\frac{2|\alpha_w|}{|\mu_-\mu_{\rm cc}|}\right)^\frac{1}{3}\,,\;\;\; \ell_\pi(\mu)\approx\left(\frac{2|\alpha_w|}{|\mu-\mu_{\rm
 sat}|}\right)^\frac{1}{3}\,,
 \ee
we see that the connection between groove evaporation and complete wetting goes much deeper. Thus, while $\ell_\pi$ diverges as $\mu\to\mu_{\rm sat}$ from below
and $\ell_E$ diverges as $\mu\to\mu_{\rm cc}$ from above, the divergences are otherwise characterised by precisely the same power-law functions, apart from the
shift in the location of the respective transition. That is
 \bb
 \ell_E(\mu-\mu_{\rm cc})=\ell_\pi(\mu_{\rm sat}-\mu)\,.
 \label{cov}
 \ee
This is a further example of a covariance relation relating adsorptions at phase transitions on different substrates, similar that observed in studies of wedge
and cone filling \cite{cov,cov2,cov3}.

We have tested this prediction using our microscopic DFT by comparing the divergences of $\ell_\pi=\Gamma/\Delta\rho$ and $\ell_E=\Gamma_E/\Delta\rho$ at a temperature $T/T_w=1.1$ in a groove of depth $D=50\sigma$ and width
$L=12\sigma$. The results are shown in Fig. 11 and show a near perfect collapse of the two curves $\ell_\pi(\mu)$ and $\ell_E(\mu)$. In drawing the shifted curve for $\ell_E$ we have treated the value of $\mu_{\rm cc}$ as a
fitting parameter which determined as $\mu_{\rm cc}^{\rm fit}=-4.0541\varepsilon$. This compares very well with the value $\mu_{\rm cc}=-4.0529\varepsilon$ obtained from the independent 1D DFT analysis for an infinite open
slit. The small difference between these values is attributable to the finite depth of our groove.

The origin of the covariance can be easily understood by simply recasting the slab model analysis in the language of an effective potential. The equilibrium film
thickness $\ell_\pi$ of a complete wetting layer, is the minimum of a binding potential $W_\pi(\ell)$ defined as the excess grand potential per unit area of a
wetting film at a planar wall which constrained to be of height $\ell$. Within the slab model analysis this is determined in standard fashion as
 \bb
 W_\pi(\ell)=(\mu_{\rm sat}-\mu)\Delta\rho\ell-\Delta\rho\int_{\ell}^\infty V^{(1)}(z)\dd z\,, \label{wpi}
 \ee
where the first term is the thermodynamic cost of having a layer of liquid which is metastable in the bulk and the second arises from the integral over the
long-ranged wall-fluid forces. This gives rise to an effective repulsion between the interfaces
 \bb
 W_\pi(\ell)=(\mu_{\rm sat}-\mu)\Delta\rho\ell+\frac{A}{\ell^2} +\cdots\,,\label{wpi2}
 \ee
where the Hamaker constant $A=-\alpha_w\Delta\rho$, which is positive. Minimization of this determines $\ell_\pi$ in agreement with Eq.~(\ref{lpi}). Now we turn
our attention to the evaporation occurring in an infinitely deep groove and and understand the structure of the analogous binding potential $\omega^{\rm
ex}(\ell)$ where we have suppressed the dependence on the wetting films at the side walls. The slab model result for this is
 \bb
 \frac{\omega^{\rm ex}(\ell)}{L}=(\mu-\mu_{\rm cc})\Delta\rho(D-\ell)+\frac{A}{(D-\ell)^2}+{\cal{O}}\left((D-\ell)^{-3}\right)\,,\label{we}
 \ee
where we have divided by the slit width so that the dimensions are the same as for $W_\pi(\ell)$. Here the first term is now the thermodynamic cost of having a
volume of capillary-gas which is metastable in the groove. This is the direct analogue of the first term in $W_\pi(\ell)$ and determines the shift in the
location of the transition from $\mu_{\rm sat}$ to $\mu_{\rm cc}$. The second term is the all important repulsion of the meniscus from the capillary opening and
can be understood as follows: consider an infinite open capillary slit exactly at $\mu=\mu_{\rm cc}$, and place a meniscus at some arbitrary position. Now
imagine slicing off the side walls at height $D-\ell$ from the meniscus and replacing this by the vapour. The change to the excess grand potential from removing
this volume of wall involves precisely the same integral over the dispersion interaction as in Eq.~(\ref{wpi}) apart from a slab of thickness $L$ which
contributes to the error term in Eq.~(\ref{we}). Minimization of $\omega^{\rm ex}(\ell)$ recovers the above expression for $\ell_E$ and hence the covariance law.

 For completeness we remark that for the condensation transition the binding potential is given by
 \bb
 \frac{\omega^{\rm ex}(\ell)}{L}=(\mu_{\rm cc}-\mu)\Delta\rho\ell+\frac{9AL}{8\ell^3}+\cdots\,.\label{wc}
 \ee
This has a very similar interpretation to the potential for evaporation with the first term representing the energy cost of having a volume of metastable
capillary liquid for $\mu<\mu_{\rm cc}$. The reason why the meniscus repulsion from the groove bottom is higher order than for the evaporation can also be
understood by dimensional analysis: imagine first that we are at coexistence in an open slit then cap the geometry by adding a slab of  solid that fits between
the side walls a distance $\ell$ below the meniscus. Since this is of finite width $L$ the integration of the intermolecular forces over this volume produces a
power law which is one order higher compared to that for wetting at a planar wall and evaporation.

\section{Conclusion}

In this paper we have used a mean-field DFT and slab model analysis to determine the order of condensation and evaporation transitions in a deep capillary groove
with long-ranged wall-fluid forces. We have shown that the condensation transition becomes first order below the wetting temperature due to the presence of
corner menisci and shown that the complementary Kelvin equation accurately describes the associated size of the metastable regime. For evaporation on the other
hand our results indicate that the transition remains continuous at all temperatures and confirm the  mean-field value of the  critical exponent $\beta_E=1/3$ .
Our analysis of evaporation  also revealed a remarkably simple covariance relation with complete wetting at a planar wall.

Our study has been entirely at mean-field level and neglects the long wavelength, interfacial, fluctuations of the meniscus, the most dominant of which arise
from those in the height of the meniscus along the groove. As discussed in \cite{evans_cc} this means that the fluctuation theory of meniscus unbinding is
analogous to that of two dimensional complete wetting but with a stiffness parameter, resisting the undulations of the meniscus, which is $\propto \gamma L$.
Thus, for continuous condensation, the mean-field power-law divergence $\ell_C\approx ((\mu_{\rm cc}-\mu)/L)^{-1/4} $ will, as $\mu\to \mu_{\rm cc}$ , eventually
cross-over to  $\ell_C \approx (L^2(\mu_{\rm cc}-\mu))^{-1/3}$ describing the true asymptotic critical behaviour (assuming the groove is macroscopically long).
However, a simple matching of these power laws shows that the size of the asymptotic regime is negligibly small since it scales as $L^{-11}$. Thus to all intents
and purposes the mean-field description of the continuous capillary condensation is exact. Similar remarks apply to continuous evaporation. The repulsive term
$\propto (D-\ell)^{-2}$ appearing in the effective potential (\ref{we}) is marginal which implies that that value of the exponent $\beta_E=1/3$ is not altered by
fluctuation effects. The only influence of these is that they slightly change the critical amplitude of the divergence of $\ell_E$ so that the mean-field result
(\ref{le}) is multiplied by a factor $1+\mathcal{O}(\beta\gamma L^2)^{-1}$. This is only significant if the evaporation occurs in the immediate vicinity of the
capillary critical point and is otherwise entirely negligible.

An important generalization of the present study is to include fully long-ranged fluid-fluid forces which decay as $\phi(r)\propto -\epsilon/r^6$.  Indeed,
within the slab model  of evaporation it is trivial to allow for such forces, at leading order, since the integrals they introduce into the analysis are
identical to those arising from the wall-fluid forces. The upshot of this is that the prediction for the meniscus height is altered to
  \bb
 \ell_E(\mu)\approx\left(\frac{2(\alpha_f-\alpha_w)}{\mu-\mu_{\rm cc}}\right)^\frac{1}{3}
 \ee
where
 \bb
 \alpha_f=-\frac{1}{3}\pi\varepsilon\sigma^3\,.
 \ee
Equivalently, the Hamaker constant appearing in the potentials (\ref{wpi2}) and (\ref{we}) is replaced by $A\propto \Delta\rho(\alpha_f-\alpha_w$). This result
tells us that the covariance law (\ref{cov}) for evaporation occurring for $T>T_w$ and complete wetting remains unchanged.  More interestingly however it
suggests that the evaporation becomes first-order  when the Hamaker constant changes sign. This occurs under two circumstances: a) at the wetting temperature
$T_w$ associated with critical (second-order) wetting transition of the side walls b) at the spinodal temperature $T_s$ associated with first-order wetting
transition of the side walls. This is defined as the temperature at which the activation barrier in the wetting binding potential $W_\pi(\ell)$ first appears.
Interestingly, these conditions are precisely the same as the slab model predictions for the order of wedge filling transitions and point to a possible deeper
connection with that phase transition \cite{wood1}. We emphasize that this prediction of a possible change in the order of the evaporation transition is entirely
consistent with the present numerical DFT study since in our current model, the short-ranged nature of the fluid-fluid forces means there is no spinodal
temperature associated with the first-order wetting transition. However some caution is needed with this prediction regarding the change in order of evaporation
since below the wetting temperature one should also consider the higher order terms in the binding potentials for which we need to go beyond the sharp kink
approximation. For example it is certainly not the case that the higher order terms in (\ref{wpi2}) and (\ref{we}) are the same which is the reason why the
covariance law (\ref{cov}) only applies above $T_w$. In addition for very narrow slits it is necessary to carefully model the opening of the groove into the bulk
reservoir since the structure of this may lead to additional interfacial pinning. This requires more study using numerical DFT rather than simple slab model
considerations.

Finally we mention that there is one phenomenon occurring in the groove geometry that we have not considered at all. This is the capillary-wetting transition
defined as the divergence in the adsorption $\Gamma_C$ as the temperature is increased towards $T\approx T_w$ along the capillary-coexistence line $\mu=\mu_{\rm
cc}$ \cite{parry_groove}. The complementary Kelvin equation tells us that macroscopically the transition must occur at $T_w$ since this is the temperature at
which the contact angle vanishes implying the end of any metastability associated with corner menisci. The reason why we have avoided discussion of this
transition is that here the mean-field character of the DFT is unreliable. According to the present DFT the capillary wetting transition is first-order and with
it is associated a capillary pre-wetting line extending off capillary-coexistence and for $T>T^*(L)$. The first-order nature of this transition can be seen from
the structure of the condensation binding potential (\ref{wc}). While the slab model parameterization does not allow for corner menisci for $T<T_w$ it does tell
us that if a single meniscus were to be formed at a distance $\ell$ above the groove bottom then it must be repelled from it (since the Hamaker constant $A>0$).
However we know that below $T_w$ the lowest free-energy configuration is due to bound corner menisci. Thus there is always a potential barrier between bound
corner menisci and a single meniscus state located above the bottom. Thus the transition must be first-order. However this mean-field reasoning is incorrect
because the capillary wetting transition must belong to the universality of two dimensional critical wetting with short-ranged forces since the
$\mathcal{O}(\ell^{-3})$ interaction of the meniscus with the wall appearing in Eq.~(\ref{wc}) is irrelevant. This implies not only that the transition is
continuous and hence that there is no capillary pre-wetting line but that the location of $T^*(L)$ is renormalized by fluctuations and occurs below its
mean-field prediction because the meniscus can tunnel out of the barrier which binds it to the corner(s). However modelling this using an effective interfacial
Hamiltonian, while taking into account the influence of long-ranged forces, is difficult because it is necessary to model two corner menisci and single meniscus
configurations. This will be the subject of future work.

\begin{acknowledgments}

\hspace*{0.01cm}

\noindent  We are grateful to C. Rasc\'on for useful discussions. A.M. acknowledges the financial support from the Czech Science Foundation, project 13-09914S. AOP acknowledges the financial support from the EPSRC UK for grant
EP/J009636/1.
\end{acknowledgments}

\appendix

\section{Slab model integrals} \label{append}

The essential ingredient in the slab model calculation of the constrained grand potential $\omega^{\rm ex}(\ell)$ given by Eq.~(\ref{omex}) is the integration of
the external potential over the volume of the gas. Per unit length of the capillary, this is given by:
 \begin{widetext}
 \begin{eqnarray}
 \frac{1}{L_\parallel}\int_{V_g}V(\rr)\dr&=&\int_{\ell}^{D}\dd z\int_{\ell_\pi}^{L-\ell_\pi}\dd x\,V(x,z)\nonumber\\
  &=&\alpha_w\int_{\ell}^{D}\dd z\int_{\ell_\pi}^{L-\ell_\pi}\dd x\left[\frac{2}{z^3}+\psi(x,z)+\psi(x,D-z)+\psi(L-x,z)+\psi(L-x,D-z)\right]\nonumber\\
 &=&\alpha_w\int_{\ell}^{D}\dd z\int_{\ell_\pi}^{L-\ell_\pi}\dd x\left[\frac{2}{z^3}+2(\psi(x,z)+\psi(x,D-z))\right]\nonumber\\
  &=&2\alpha_w\int_{\ell}^{D}\dd z\left[\frac{L-2\ell_\pi}{z^3}+\Psi(L-\ell_\pi,z)-\Psi(\ell_\pi,z)+\Psi(L-\ell_\pi,D-z)-\Psi(\ell_\pi,D-z)\right]\,,\nonumber
 \end{eqnarray}
 \end{widetext}
 where we have defined
  \bb
 \Psi(x,z)\equiv\int\psi(x,z)\dd x=-\frac{x}{z^3}+\frac{(2x^2-z^2)\sqrt{x^2+z^2}}{2z^3x^2}\,.
 \ee
Thus, when we minimise $\Omega^{\rm ex}$ with respect to $\ell$, we can make use of the relation
 \begin{widetext}
 \bb
 \frac{\partial}{\partial
\ell}\left[\frac{1}{L_\parallel}\int_{V_g}V(\rr)\dr\right]=-2\alpha_w\left[\frac{L-2\ell_\pi}{\ell^3}+\Psi(L-\ell_\pi,\ell)-\Psi(\ell_\pi,\ell)
+\Psi(L-\ell_\pi,D-\ell)-\Psi(\ell_\pi,D-\ell)\right]\,, \label{omlz}
 \ee
 \end{widetext}
 which can be expanded
\begin{widetext}
 \bb
 \frac{\partial}{\partial
\ell}\left[\frac{1}{L_\parallel}\int_{V_g}V(\rr)\dr\right]=-2\alpha_w\left[\frac{1}{\ell_\pi^2}-\frac{1}{(L-\ell_\pi)^2}+
\frac{2\ell_\pi-L}{(D-\ell)^3}+\frac{9}{16}\frac{L(L-2\ell_\pi)}{\ell^4}+\cdots\right]\,. \label{omlz2}
 \ee
 \end{widetext}
Combining with (\ref{omex}) one obtains
\begin{widetext}
 \bb
 (p-p_l^+ )(L-2\ell_\pi) -2\gamma+2\alpha_w\Delta\rho\left[\frac{1}{\ell_\pi^2}-\frac{1}{(L-\ell_\pi)^2}+
\frac{2\ell_\pi-L}{(D-\ell)^3}+\frac{9}{16}\frac{L(L-2\ell_\pi)}{\ell^4}+\cdots\right]=0
 \ee
 \end{widetext}
 and substituting $\ell_\pi$ from Eq.~(\ref{lpi}) gives:
 \begin{widetext}
 \bb
 (p-p_l^+ )(L-3\ell_\pi) -2\gamma+2\alpha_w\Delta\rho\left[
\frac{2\ell_\pi-L}{(D-\ell)^3}+\frac{9}{16}\frac{L(L-2\ell_\pi)}{\ell^4}+\cdots\right]=0\,.
 \ee
 \end{widetext}
which determines the equilibrium height of the meniscus in a finite depth groove. Finally, using $p-p_l^+\approx (\mu_{\rm sat}-\mu)\Delta\rho$ and
Eq.~(\ref{kelvin}) we find that for $\mu<\mu_{\rm cc}$, in the limit $D\to \infty$, the equilibrium position of the meniscus height above the groove bottom
$\ell_C$ satisfies
  \begin{eqnarray}
\mu&=&\mu_{\rm cc}(L)+\frac{9}{8}\frac{\alpha_w(L(L-2\ell_\pi))}{(L-3\ell_\pi)}\frac{1}{\ell_C^4}+\cdots \nonumber\\ 
&\approx& \mu_{\rm cc}(L)+\frac{9}{8}\frac{\alpha_wL}{\ell_C^4}
  \end{eqnarray}
Conversely, for  $\mu>\mu_{\rm cc}(L)$, in the limit $D\to \infty$, with $D-\ell$ fixed, the equilibrium distance of the meniscus from the top of the groove
$\ell_E$ satisfies
  \begin{eqnarray}
\mu&=&\mu_{\rm cc}(L)-\frac{2\alpha_w(L-2\ell_\pi)}{(L-3\ell_\pi)}\frac{1}{\ell_E^3}+\cdots \nonumber\\ 
&\approx&\mu_{\rm cc}(L)-\frac{2\alpha_w}{\ell_E^3}\,.
  \end{eqnarray}

\end{document}